\documentclass[prd,aps,reprint,11pt,onecolumn]{revtex4-1}

\usepackage{amsfonts,amsmath,amssymb,graphicx,bm,subfigure,tensor,multirow}
\usepackage[colorlinks,
  linkbordercolor={0 0 0},
  pdfborder={0 0 0},
  linkcolor=blue,
  citecolor=blue,
  urlcolor=blue,
  breaklinks=true]{hyperref}

\def\eqarray#1{\begin{eqnarray} #1 \end{eqnarray}}
\newlength{\x}
\newlength{\z}
\newlength{\abc}
\begin{document}

\author{Mridupawan Deka$^{1)}$}
\email{mpdeka@theor.jinr.ru}

\author{Maxim Dvornikov$^{1), 2)}$}
\email{maxim.dvornikov@gmail.com}

\title{The effect of background matter on the spin oscillations of neutrinos scattered
by the supermassive black hole}

\affiliation{$^{1)}$\,Bogoliubov Laboratory of Theoretical Physics, Joint Institute for Nuclear Research, Dubna, Russia; \\
$^{2)}$\,Pushkov Institute of Terrestrial Magnetism, Ionosphere and Radiowave Propagation (IZMIRAN), Troitsk, Moscow, Russia}

\begin{abstract}
    
  We study spin oscillations of neutrinos in relativistic moving matter inside
  an accretion disk. These neutrinos are gravitationally scattered off
  a spinning Kerr black hole surrounded by a thick accretion disk. The disk can co-rotate
  and counter-rotate with respect to BH spin. We perform numerical simulations of
  the propagation of a large number of incoming test neutrinos. We briefly discuss
  our results.
\end{abstract}

\maketitle

\section{Introduction}
\label{sec:INTR}

Since neutrinos are experimentally confirmed to have non-zero masses as well as
undergo mixing between flavors, it results in flavor oscillations (see,
e.g., refs.~\cite{NOvA:2021nfi,GiuKim07}), it provides a  unique tool to explore
the physics beyond the standard model.

In addition to photons, it is found that a significant flux of ultra-relativistic
neutrinos are also emitted by the accretion disk of a black hole
(BH)~\cite{Caballero:2011dw}. These neutrinos move in strong gravitational fields in
the vicinity of the BH. Consequently, they experience electroweak interactions with
the background matter of the disk~\cite{Okun:1986na}. Some of the recent works,
e.g.Refs.~\cite{Pustoshny:2018jxb,Studenikin:2004bu}, discuss the effect of the
relativistic motion of the background matter on neutrino spin oscillations. It is
shown that the interactions with the transversal matter currents can give rise to spin
oscillations within the same flavor.

Due to the spin oscillations, the left polarized active neutrinos become sterile or
right handed ones, meaning they can not be observed. If we consider only the
gravitationally scattered neutrinos, their ``in'' and ``out'' states are in the
asymptotically flat spacetime. Therefore, their spin states are well defined.
In the event of spin oscillations, we shall see a reduction in the initial flux of
neutrinos at the observer position since the right handed neutrinos can not detected.

However, the Lorentz factor, $\gamma \rightarrow \infty$, in flat spacetime in the case
of ultra-relativistic neutrinos. It can be argued that the interaction due to the
transversal matter current is negligible since it is inversely proportional to $\gamma$.
However, one cannot guarantee that the nontrivial matter motion in an accretion disk,
driven by a strong gravitational field in the vicinity of BH, does not induce a spin-flip
even for neutrinos which are ultrarelativistic in asymptotically flat spacetime far away
from BH. We can study this phenomena only numerically and see whether we observe any spin
oscillation due to the transversal background matter only.

For this study, we follow the same procedure for gravitationally scattered neutrinos
by a rotating BH as in the previous studies, e.g.~\cite{Dvo06, Dvo13, Dvo23c,Dvo23d,
  Dvo23a, Dvo23b,Deka:2023ljj,Deka:2025eub,Deka:2025war}, except that this time 
we do not consider any magnetic field inside the accretion disk. However,
we still use the same ``Polish doughnut'' model for the accretion
disk~\cite{Abramowicz_1978}. Similar to~\cite{Deka:2025eub}, we consider that
the incoming neutrino beam traversing at a random angle with respect to BH spin.
We also consider both the situations where the disk co-rotates and counter-rotates
with respect to BH spin.

This work is organized in the following way. First, in Sec.~\ref{sec:formalism},
we describe our approach. In Secs.~\ref{sec:MOTION} and~\ref{sec:SPIN_EV},
we discuss the motion in the gravitational field of a rotating BH and spin evolution
of a test particle along its trajectory, respectively. Also, the
interactions with background matter are discussed. In Sec.~\ref{sec:accretion_disk}, we
describe the accretion disk used in this work, and the numerical parameters are
described in Sec.~\ref{sec:NUMERICAL}. The results and conclusions are presented in
Sec.~\ref{sec:RES}.

\section{A Brief Description of our study}
\label{sec:formalism}

\begin{figure}[htbp]
\centering
\subfigure[]
 {\label{fig:tourus_2}
   \includegraphics[width=0.47\hsize]{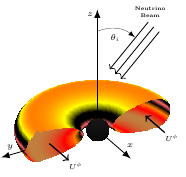}
 }
 \hspace{0mm}
 \subfigure[]
 {\label{fig:tourus_4}
   \includegraphics[width=0.47\hsize]{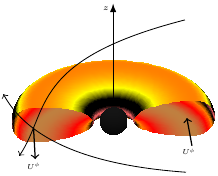}
 }
\protect 
\caption{Schematic diagrams showing neutrino trajectories.
  (a) A beam of neutrinos coming from $\infty$ at an angle $\theta_i$ with
  respect to the BH spin.
  (b) Neutrinos crossing the equatorial plane through the accretion disk.
  The transeversal matter velocity is perpendicular to their momenta.}
\end{figure}

We consider that a beam of left-handed Dirac neutrinos, emitted from a distant
source, approaches a spinning SMBH with a polar angle $\theta_i$ with respect
to the SMBH spin (See Fig.~\ref{fig:tourus_2}). We can write their initial
co-ordinates as $(r,\theta,\phi)_{\mathrm{source}} = (\infty,\theta_i,0)$. The
neutrinos, which are not captured by the BH, are gravitationally scattered.
In this study, we are interested only in the scattered neutrinos.

From Fig~\ref{fig:tourus_4}, we see that some of the scattered neutrinos pass
through the equatorial plane of the BH through the accretion disk. In such a
situation, they undergo interactions with the transverse component of the matter
currents. As discussed in~\cite{Pustoshny:2018jxb,Studenikin:2004bu}, this may
result in spin oscillations. As a result, some of the left-handed
neutrinos of the initial beam become right-handed or sterile.

Since the sterile neutrinos can not be observed, there will be a reduction in the
flux of the initial beam of neutrinos at the observer position,
$(r,\theta,\phi)_{\mathrm{obs}} = (\infty,\theta_\mathrm{obs},\phi_\mathrm{obs})$.
Our goal is to study the probability distributions of the handedness of the
observed neutrinos as functions of $\theta_\mathrm{obs}$ and $\phi_\mathrm{obs}$.
In the event of no spin oscillation, we shall find the probability to be $1$
everywhere in the $\theta_\mathrm{obs}$ and $\phi_\mathrm{obs}$ plane.

In the following, we discuss the trajectories of the neutrinos, their spin
evolution along them as well as some discussions of the accretion disk.

\section{The curved Spacetime of a spinning Black Hole and Ultra-relativistic neutrinos}
\label{sec:MOTION}

The scattered neutrinos move through the curved spacetime of a rotating BH
before being observed at the observer position at the asymptotically flat space-time.
We describe such curved spacetime in Kerr metric which can be written
in Boyer-Lindquist coordinates, $x^{\mu}=(t,r,\theta,\phi)$, as,
\begin{equation}\label{eq:Kerrmetr}
  \mathrm{d}s^{2}=g_{\mu\nu}\mathrm{d}x^{\mu}\mathrm{d}x^{\nu}=
  \left(
    1-\frac{rr_{g}}{\Sigma}
  \right)
  \mathrm{d}t^{2}+2\frac{rr_{g}a\sin^{2}\theta}{\Sigma}\mathrm{d}t\mathrm{d}\phi-\frac{\Sigma}{\Delta}\mathrm{d}r^{2}-
  \Sigma\mathrm{d}\theta^{2}-\frac{\Xi}{\Sigma}\sin^{2}\theta\mathrm{d}\phi^{2},
  \notag
\end{equation}
where $M$ is the BH mass, and $J$ is the angular momentum along the $z$-axis so that
 $J = a M\, (0 < a < M)$. $r_g = 2M$ is the Schwarzschild radius, and, 
\begin{equation}\label{eq:dsxi}
  \Delta=r^{2}-rr_{g}+a^{2},
  \quad
  \Sigma=r^{2}+a^{2}\cos^{2}\theta,
  \quad
  \Xi=
  \left(
    r^{2}+a^{2}
  \right)
  \Sigma+rr_{g}a^{2}\sin^{2}\theta. \notag
\end{equation}
The neutrino energy ($E$), its angular momentum ($L$) and the Carter constant ($Q$) are
constants of motion. Note that $Q>0$ in case of scattering. We use the variables,
$r = xr_{g},\, L = yr_{g}E,\, Q = wr_{g}^{2}E^{2},\, a = zr_{g},\, t = \cos\theta$,
to make all relevant quantities dimensionless.

In this curved spacetime, we can exactly describe the motion of an ultra-relativistic
neutrino~\cite{GraLupStr18}. We extensively discuss it in some of our previous
works, e.g.~\cite{Dvo23b,Deka:2023ljj,Deka:2025eub,Deka:2025war}. We use the
relationships obtained there in this work also.

\section{The evolution of Neutrino polarization}
\label{sec:SPIN_EV}

A neutrino interacts electroweakly with the accretion disk. In this section, we discuss
the evolution of neutrino polarization along its trajectory.

The polarization of a neutrino is described by an invariant three spin vector
$\bm{\zeta}$ in the rest frame with respect to a locally Minkowskian frame.
The evolution of the neutrino polarization vector obeys,
\begin{equation}\label{eq:nuspinrot}
  \frac{\mathrm{d}\bm{\bm{\zeta}}}{\mathrm{d}t}=2(\bm{\bm{\zeta}}\times\bm{\bm{\Omega}}),
\end{equation}
where,
\begin{equation}
  {\bm{\Omega}}={\bm{\Omega}}^{\mathrm{g}} + {\bm{\Omega}}^{\mathrm{matt}}. \notag
\end{equation}
The discussion of the gravitational interactions, ${\bm{\Omega}}^{\mathrm{g}}$,
is provided in Refs.~\cite{Dvo23a,Dvo23b,Deka:2025eub}.

On the other hand, it is shown in~\cite{Pustoshny:2018jxb,Studenikin:2004bu}
that the matter interactions can be decomposed into transversal and
longitudinal parts such that,
\eqarray{
  {\bm{\Omega}}^{\mathrm{matt}}
  &=& \frac{1}{\gamma}
  ({\bm{\Omega}}^{\mathrm{matt}}_\perp + {\bm{\Omega}}^{\mathrm{matt}}_\parallel)
}
where, $\gamma = (1 - \beta^2)^{-1/2}$, $\bm{\beta}$ being the neutrino velocity
provided that a neutrino moves in the flat spacetime. The explicit forms of
${\bm{\Omega}}^{\mathrm{matt}}_\perp$ and ${\bm{\Omega}}^{\mathrm{matt}}_\parallel$ are given
in~\cite{Pustoshny:2018jxb,Studenikin:2004bu}.

It is shown in~\cite{Pustoshny:2018jxb,Studenikin:2004bu} that
in the presence of a non-zero transverse component,
${\bm{\Omega}}^{\mathrm{matt}}_\perp$, there is a finite probability that the
spin oscillations do occur. Since we are dealing with ultra-relativistic neutrinos,
$\gamma \rightarrow \infty$ at the source and observer positions which are
located in the asymptotically flat spacetime. Since ${\bm{\Omega}}^{\mathrm{matt}}_\perp$ is
inversely proportional to $\gamma$, it is, in fact, zero in the case of ultra-relativistic
neutrinos resulting in no spin oscillations.

We mention that the analogue of the Lorentz factor, as the function of the neutrino
velocity, is not well defined in the curved spacetime near a spinning BH since we are
dealing with two reference frames in describing neutrino oscillations. One of them is
related to the world coordinates $x^\mu$ in Eq.~\eqref{eq:Kerrmetr}. Another reference
frame is based on the locally Minkowskian coordinates $x^a = e^a_{\,\mu} x^\mu$, where
$e^a_{\,\mu}$ are the vierbein vectors. Thus, the quantity
${\bm{\Omega}}^{\mathrm{matt}}_\perp / \gamma$ can become non-zero.
If spin oscillations occur in such a situation, it can be observed at the observer
position. Owing to the highly non-trivial nature, We can study this phenomena only
numerically.
     
Instead of the Eq.~\eqref{eq:nuspinrot}, it is more numerically convenient
to study the effective Schr\"odinger equation for the
neutrino polarization,
\begin{equation}\label{eq:Schreq}
  \mathrm{i}\frac{\mathrm{d}\psi}{\mathrm{d}x}= \hat{H}_{x}\psi,
\end{equation}
where,
\begin{equation}
  \hat{H}_{x}= -\mathcal{U}_{2}(\bm{\bm{\sigma}}\cdot\bm{\bm{\Omega}}_{x})\mathcal{U}_{2}^{\dagger},
  \quad
  \bm{\bm{\Omega}}_{x} =  r_{g}\bm{\bm{\Omega}}\frac{\mathrm{d}t}{\mathrm{d}r},
  \quad
  \mathcal{U}_{2}=\exp(\mathrm{i}\pi\sigma_{2}/4).\notag
\end{equation}
Here $\bm{\bm{\sigma}}=(\sigma_{1},\sigma_{2},\sigma_{3})$ are the Pauli matrices.
The Hamiltonian $\hat{H}_{x}$ is the function of $x\, ({\rm{or}}, r)$ and $\theta$.

We solve the Eq.~\eqref{eq:Schreq} numerically at every $x$. Since all incoming
neutrinos are considered to be left polarized at the infinity, the
initial condition can be written in the form, $\psi_{-\infty}^{\mathrm{T}}=(1,0)$.
We use discrete and irregular grid for the motion of the neutrinos
having denser points gradually towards the BH. We use $4$-th order
Adam-Bashforth-Moulton predictor-corrector iterative method with
the appropriate convergence conditions given as,
\begin{eqnarray}
  \left|\frac{\psi^j(x)_k -  \psi^j(x)_{k-1}}
             {\psi^j(x)_{k-1}}\right| \leq 10^{-15},
             \hspace{2mm}
             \left|\psi(x)_k\right|^2 = 1 \pm O(10^{-15}),
             \notag
\end{eqnarray}
where, $k$ is the iteration number in the determination of
the $j$-th component of $\psi(x)$. The first equation deals
with the convergence of each of the four components of $\psi(x)$,
and the second one checks the normalization condition.

After finding
$\psi_{+\infty}^{\mathrm{T}}=(\psi_{+\infty}^{(\mathrm{R})},\psi_{+\infty}^{(\mathrm{L})})$
numerically, we then compute
$P_{\mathrm{LL}} =|\psi_{+\infty}^{(\mathrm{L})}|^{2}$ at the observer
position. Here, we take into account that the neutrino velocity
changes its direction in the locally Minkowskian frame after the scattering.
If spin oscillations occur, $P_{\mathrm{LL}}$ shall be less than $1$. 
If there is no spin oscillation, $P_{\mathrm{LL}} = 1$.

\section{``Polish doughnut'' model of Accretion Disk}
\label{sec:accretion_disk}

There are several models to describe the accretion disk surrounding a
BH~\cite{Abramowicz:2011xu}. In our study, it is desirable that
a neutrino path inside the disk is long enough such that the spin rotates
a sizable angle with respect to the neutrino velocity. Therefore, 
we choose a thick ``Polish doughnut'' disk model proposed
in Ref.~\cite{Abramowicz_1978}.

We assume that the accretion disk can rotate around the BH with relativistic
velocity. We also assume that the hydrogen plasma in the accretion disk is
unpolarized and electrically neutral, i.e. the invariant number density of electrons
and protons are equal, $n_e = n_p$. We treat the electroweak interaction of a
neutrino with the fermions in the plasma in the forward scattering
approximation~\cite{DvoStu02}. The four potential, $G^\mu$, for the neutrino electroweak
interactions with a background matter can then be written as,
\begin{eqnarray}
  \label{eq:Gmu_2}
  G^\mu &=& \displaystyle\sum_{f = e,p}  q_f J_f^\mu,
\end{eqnarray}
where  $J_f^\mu = n_f U_f^\mu$ are the hydrodynamic currents, and $U_f^\mu$ are
the four velocities of fermions in the disk. We assume that $U_e^\mu = U_p^\mu$,
i.e. there is no differential rotation between the components of the plasma.

The only non-zero components in Eq.~\eqref{eq:Gmu_2} are $J_f^t = n_f U_f^t \neq 0$
and $J_f^\phi = n_f U_f^\phi \neq 0$ because of the axial symmetry of the disk.
As a result, all the parameters of the disk depend on $r$ and $\theta$.

We also assume that the specific angular momentum of a particle in the disk
$l = L/E$ is constant, $l=l_{0}$. Here $E = m U_{f}^{t}$ is the energy of the
particle, $L = - m U_{f}^{\phi}$ is its angular momentum, and $m$ is its mass.

If we define the generating function for the disk potential
as~\cite{Abramowicz_1978,Kom06},
\eqarray{
  \label{eq:genfunW}
  W(r,\theta) &=&
  \frac{1}{2}\ln
  \left|
  \frac{\mathcal{L}}{\mathcal{A}}
  \right|, 
}
then disk density, $\rho$, can be written as,
\eqarray{
  \label{eq:rhopm}
  \rho &=&
  \left[
    \frac{\kappa-1}{\kappa}
    \frac{W_{\mathrm{in}}-W}{K}
    \right]^{\frac{1}{\kappa-1}},
}
where $K$ and $\kappa$ are the constants in the equations of state.  For computation,
we use the dimensionless variables $\tilde{K}=r_{g}^{4(1-\kappa)}K$. We choose
$\kappa=4/3$~\cite{Kom06}. The parameter $W_{\mathrm{in}}$ is the value of $W$ at the
inner border of the disk. We set its value, $W_{\mathrm{in}} = 10^{-5}$.

The components of $U_{f}^{\mu}$ can then be written as,
\eqarray{
  \label{eq:UB}
  U_{f}^{t} &=&
  \sqrt{
  \left|
  \frac{\mathcal{A}}{\mathcal{L}}
  \right|
  }
  \frac{1}{1-l_{0}\Omega},
  \hspace{2mm}
  U_{f}^{\phi} \,=\, \Omega U_{f}^{t},
}
where,
\eqarray{
  \label{eq:Omegadisk}
  \mathcal{L} &=& g_{tt}g_{\phi\phi}-g_{t\phi}^{2},\hspace{2mm}
  \mathcal{A}\, =\, g_{\phi\phi}+2l_{0}g_{t\phi}+l_{0}^{2}g_{tt},\hspace{2mm}
  \Omega\, =\, -\frac{g_{t\phi}+l_{0}g_{tt}}{g_{\phi\phi}+l_{0}g_{t\phi}}.
}
Here, $\Omega$ is the angular velocity in the disk.

Equations~(\ref{eq:genfunW})-(\ref{eq:Omegadisk}) fully define all
the characteristics of the disk. The methods to determine the values of
$\tilde{K}$ and $\rho$ for various BH spin are discussed in Sec.~\ref{sec:NUMERICAL}

The value of the specific angular momentum, $l_0$, varies according to whether the disk
is co-rotating or counter-rotating with BH spin~\cite{Bardeen:1972fi}. In the current work,
we consider both the co-rotating or counter-rotating disks. Both the cases are discussed
in details in~\cite{Deka:2025eub}.

\begin{table}[hbtp]
  \caption{The values of the parameters that are common to
    various combinations of BH spins and incident angles.}
  \label{tab:num_para_1}
  \begin{center}\setlength{\tabcolsep}{8pt}
    \begin{tabular}{c|c|c|c}
      \hline\hline
      {\bf BH mass}
      & {\boldmath $W_{\mathrm{in}}$}
      & {\boldmath $n_e^{\mathrm{max}}$}
      & {\boldmath $r_g$}\\
      \hline\hline
      $10^8 M_\odot$
      & $10^{-5}$
      & $10^{18}\,\text{cm}^{-3}$
      & $3\times 10^{13}$ cm\\
      \hline\hline
    \end{tabular}
  \end{center}
\end{table}

\begin{table}[hbtp]
  \caption{The values of $\tilde{K}$ for
    different BH spins for both the co-rotating and
    counter-rotating disks.}    
  \label{tab:num_para_2}
  \begin{center}\setlength{\tabcolsep}{8pt}
    \begin{tabular}{c|c|c}
      \hline\hline
      {\bf Disk type}
      & {\bf BH Spin}
      & {\boldmath $\tilde{K}$}\\
      \hline\hline
      \multirow{3}{*}{\bf Co-rotating}
      & {$a = 0.02M$}
      & $2.51\times 10^{-31}$\\
      \cline{2-3}
      & {$ a = 0.50M$}
      & $3.53\times 10^{-31}$\\
      \cline{2-3}
      & {$ a = 0.98M$}
      & $9.55\times 10^{-31}$\\
      \hline
      \multirow{3}{*}{\bf Counter-rotating}
      & {$a = 0.02M$}
      & $2.43\times 10^{-31}$\\
      \cline{2-3}
      & {$ a = 0.50 M$}
      & $1.95\times 10^{-31}$\\
      \cline{2-3}
      & {$ a = 0.98M$}
      & $1.65\times 10^{-31}$\\
      \hline\hline
    \end{tabular}
  \end{center}
\end{table}

\section{Numerical Parameters}
\label{sec:NUMERICAL}

The SMBH is surrounded by a thick accretion disk which consists of a hydrogen plasma.
It either co-rotates or counter-rotates around the SMBH
with a relativistic velocity. We consider  both the cases. 

We consider three BH spins,
$a\,=\,2\times 10^{-2} M, 0.50 M$ and $0.98 M$. For each BH spin, we consider
eight different incident angles,
$\cos\theta_i = \pm 0.01, \pm 0.50. \pm 0.707$ and $\pm 0.90$.
We fix the mass of the SMBH at $M = 10^8 M_\odot$. The number of neutrinos we use
in each combination of BH spin and incident angle $\theta_i$ is more than $2$ million
for both co-rotating and counter-rotating disks.

The values of the parameters that are common to various BH spins and
incident angles are listed in Table~\ref{tab:num_para_1}.

In order to determine the parameters $\tilde{K}$, we choose the  maximal
number density of electrons to be
$n_e^{\mathrm{max}} = 10^{18}\,\text{cm}^{-3}$~\cite{Jia19} such that the
normalized electron number density distributions is $n_{e}/10^{18}\,\text{cm}^{-3}$,
where $n_{e}=\rho/m_{p}$, $m_p$ being the mass of proton. This provides us
the value of $\rho^{\mathrm{max}}$.  Using Eq.~\eqref{eq:rhopm}, we vary
$\tilde{K}$ to match to $\rho^{\mathrm{max}}$ for each spin of BH. Note that the value
of the generating function, $W$, also depends whether the disk is co-rotating or
counter-rotating. Therefore, one has to find the values of
$\tilde{K}$ for both the co-rotating and counter-rotating disks for each BH spin.
The values of $\tilde{K}$ are listed in Table~\ref{tab:num_para_2}.

\section{Results and Conclusion}
\label{sec:RES}

We study the case where only background matter from the accretion disk is present
without introducing any magnetic field in the disk. The purpose is to numerically check
for any evidence of spin oscillations in the presence of only gravity and background matter
as discussed in~\cite{Pustoshny:2018jxb,Studenikin:2004bu}. In our simulations,
we find that $P_{\mathrm{LL}} = 1$ for every incident angle with all cases of BH spin
for both co-rotating and counter-rotating disks. Therefore, we do not observe any spin
oscillation in the presence of gravity and background matter.

\begin{acknowledgments}
  We thank A. F. Zakharov for useful discussion.All our numerical computations have
  been performed at Govorun super-cluster at Joint Institute for Nuclear Research, Dubna.
\end{acknowledgments}

\end{document}